\documentclass[aps,twocolumn,graphicx]{revtex4-1}
\usepackage{graphicx}

\draft 

\begin{document}


\title{Reproducible synthetic method for the topological superconductor Cu$_x$Bi$_2$Se$_3$} 



\author{Ryusuke Kondo$^1$\thanks{E-mail:Kondo@science.okayama-u.ac.jp}, Taiki Yoshinaka$^2$, Yoshinori Imai$^2$, and Atsutaka Maeda$^2$}
\affiliation{$^1$Department of Physics, Okayama University, 3-1-1 Tsushimanaka, Okayama 700-8530, Japan\\
$^2$Department of Basic Science, University of Tokyo, 3-8-1 Komaba, Meguro-ku, Tokyo 153-8902, Japan\\}


\date{\today}

\begin{abstract}
We report a reproducible synthetic method for superconducting Cu-intercalated Bi$_2$Se$_3$ by an improved melt growth method. Avoiding the production of Cu$_2$Se, which has a higher melting point than that of Bi$_2$Se$_3$, and quenching Cu-Bi-Se mixtures at the liquid phase are keys to obtaining good superconducting samples in a reproducible manner.
\end{abstract}

\pacs{}

\maketitle 

Recently, a new class of materials, named as a topological insulator, has been predicted theoretically.\cite{r0,r1,r2} This type of insulator is characterized by a bulk band gap and topologically-protected metallic surface states in the band gap. Subsequently, surface sensitive experiments such as Angle-resolved photoemission spectroscopy (ARPES) \cite{r3,r4,r4a} and Scanning tunneling spectroscopy (STS) \cite{r5,r6} showed the existence of the surface state in the bulk band gap of Bi-based materials such as Bi$_{1-x}$Sb$_x$, Bi$_2$Se$_3$, and Bi$_2$Te$_3$, and confirmed that they compose the material group of topological insulator.

One of the next research directions in the field of the topological order of materials is the subject of topological superconductors.\cite{r6k,r7,r8,r12,r13,r2} The concept of the topological superconductor is a direct analogy between topological band theory and superconductivity, where the superconducting gap corresponds to the gap of the band insulator. It is characterized by a superconducting gap in the bulk, and topologically-protected gapless state in the superconducting gap. The topological superconductors have attracted much attention since they are predicted to have Majorana Fermions as protected bound states on the edge. Indeed, Majorana Fermion is a strange excitation, where particle is equal to anti-particle, and can be utilized as a candidate of novel type of quantum computing.\cite{r9}  Therefore, it is extremely important to understand physical properties of topological superconductors both in the ground state and in the excited state. 

Hor {\it et al.} reported superconductivity in Cu-intercalated Bi$_2$Se$_3$, Cu$_x$Bi$_2$Se$_3$ with 0.1$\leq${\it x}$\leq$0.3.(the optimal single crystal compositions with 0.12$\leq${\it x}$\leq$0.15)\cite{r10}  Cu atoms intercalated into the van der Waals gaps between Bi$_2$Se$_3$ layers act as electron dopants, leading to superconductivity at around 3.8 K.  Since Bi$_2$Se$_3$ is one of typical Bi-based topological insulators and the topological surface states remained in the band gap originating from the host material, Bi$_2$Se$_3$,\cite{r11} Cu$_x$Bi$_2$Se$_3$ was expected to be the first example of the topological superconductor. Thus, it is urgent task to study the physical properties of Cu$_x$Bi$_2$Se$_3$.@

However, it has been known that it is difficult to obtain high-quality samples of Cu$_x$Bi$_2$Se$_3$.  In fact, superconducting shielding fractions firstly reported by Hor {\it et al.} were rather small ($\sim$ 20 $\%$) in comparison with other superconducting materials and the resistance did not reach zero below Tc.

Kriener {\it et al.} electrochemically synthesized high-quality samples of Cu$_x$Bi$_2$Se$_3$\cite{r14}, the SC volume fraction of which was about 50 $\%$. In the superconducting state of their sample, the tunneling spectra with zero-bias conductance peaks was observed by point contact spectroscopy,\cite{r16} suggesting the existence of the gap-less state in the superconducting gap, consistent with a theoretical predictions for Majorana excitations, and this is considered to be an evidence for the realization of the topological superconductivity. On the other hand, a more recent tunneling study\cite{r17} and an Andreev reflection study\cite{r18} did not find any zero-bias anomaly. Therefore, further extensive investigations are necessary to establish the existence of topological superconductivity in these materials. 

However, it is still difficult to synthesize superconducting samples, even using the method reported by Kriener {\it et al.} This means we still need to look for different methods to fabricate good superconducting sample of Cu$_x$Bi$_2$Se$_3$ in a reproducible manner.

In this paper, we develop a new method to synthesize good superconducting Cu$_x$Bi$_2$Se$_3$ samples based on the discussion of the reason why the conventional melt growth method could not achieve good results. We confirmed that Cu atoms were consumed by the production of the by-product, Cu$_2$Se, in the stoichiometric mixture, which avoided Cu-intercalated Bi$_2$Se$_3$ to be synthesized. Cu$_2$Se is produced by the direct reaction of Cu and Se sublimated in the evacuated sealed ampule in the process of the increase in temperature in a furnace. Since the melting point of Cu$_2$Se is 1050 $^\circ$C, Cu$_2$Se produced in the initial process seems not to be melting during the process of keeping the mixture in the quartz tube overnight at 850 $^\circ$C, which exists also in Hor's procedure.  Therefore, the most essential point is to avoid the production of Cu$_2$Se during the initial reaction process.  Based on this idea, we developed an improved melt growth method, the so-called two-step method, where we use CuSe, higher oxidized Cu-Se material, as a precursor in order to avoid the production of Cu$_2$Se in the crystal growth of Cu$_x$Bi$_2$Se$_3$. By this improvements, and also by the refinement of the quenching condition, we become always able to obtain samples, which mostly attain zero resistances and found that SC is observed in Cu$_x$Bi$_2$Se$_3$ for an extraordinary wide {\it x} range, 0.03$\leq$x$\leq$0.5.

\begin{figure}
\includegraphics[clip,width=8.0cm]{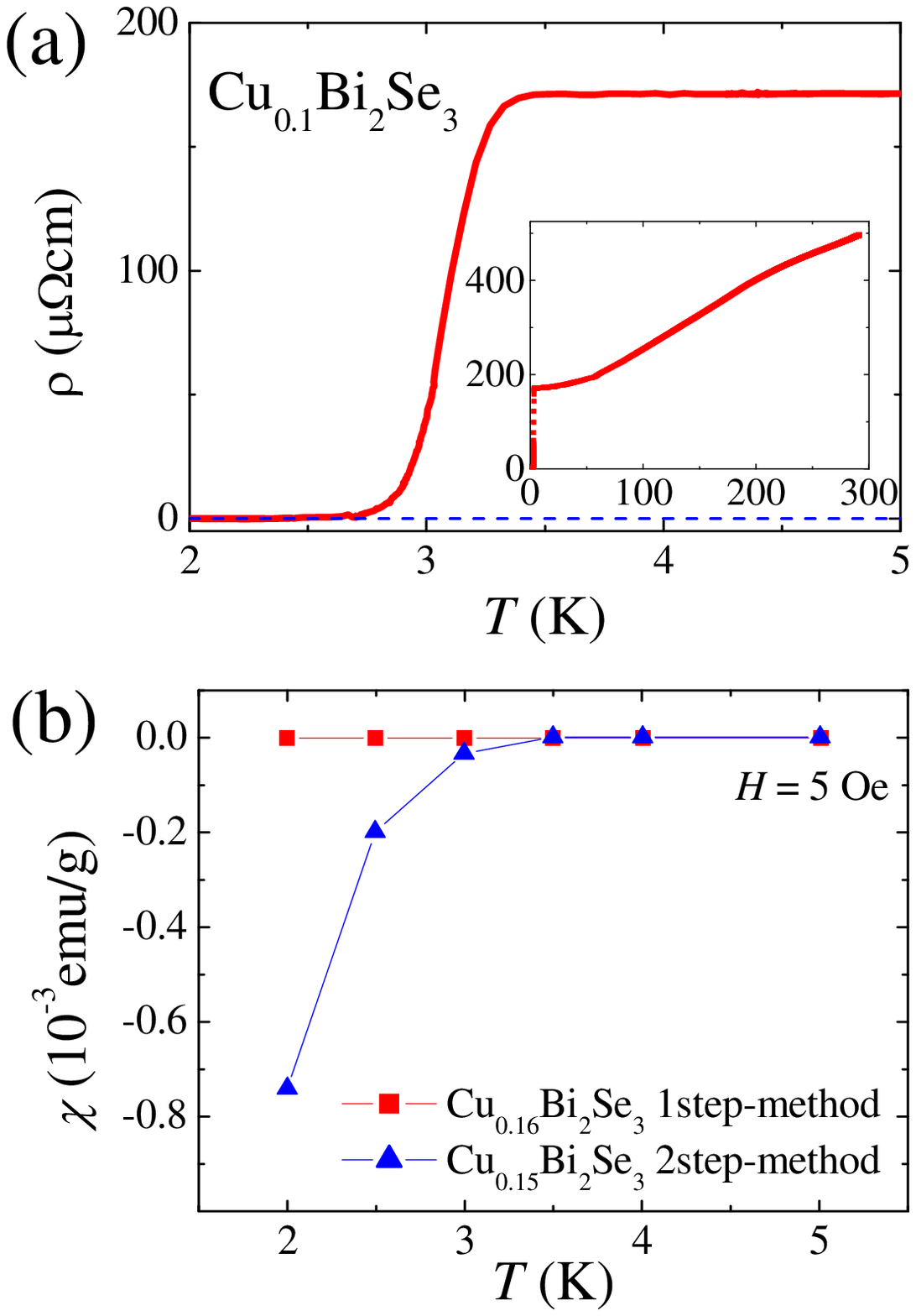}
\caption{(a) Temperature dependence of resistivity showing a zero resistance at T$_c$=2.7 K. The inset shows the temperature dependence of the resistivity for wide temperature region. (b) The temperature dependence of the magnetization of samples synthesized by the two-step method and the single step method. Magnetization data of the sample fabricated by the two step method shows the SC volume fraction of 15 $\%$ at 2.0 K.}
\label{f1}
\end{figure}

In the two-step method, we first synthesize CuSe as a precursor, since the melting point of CuSe (387 $^\circ$C) is much lower than that of Bi$_2$Se$_3$, 710 $^\circ$C; The final stoichiometric mixtures of Cu and Se ({\it i. e.} 0.15:3 in Cu$_{0.15}$Bi$_2$Se$_3$) are sealed in an evacuated quartz tube, and are kept at around 900 $^\circ$C overnight. Then, it is quenched in cold water to avoid the production of Cu$_2$Se. According to the equilibrium phase diagram of Cu-Se system,\cite{r19} Cu$_2$Se produces widely for all value of Cu/Se ratio.  Thus, the quenching process is always needed to avoid the production of Cu$_2$Se. The absence of Cu$_2$Se, and the presence of CuSe are confirmed by XRD measurements. After the above procedure, the quartz tube is opened and the stoichiometric quantity of Bi is added. Hereafter Hor's procedures are applied. 

The superconducting samples were characterized by measurements of the dc magnetization and transport properties. A commercial SQUID magnetometer (Quantum Design, MPMS5) was used to obtain the magnetization data. The electrical resistivity and the Hall coefficient R$_H$ were measured by a standard six-probe technique, where the electrical current was applied in the {\it ab} plane, and magnetic field was applied perpendicular to the {\it ab} plane. 

Figure $\ref{f1}$ shows the temperature dependence of (a) electrical resistivity and (b) magnetic susceptibility of a typical sample (x=0.1) obtained by the two-step method at around the superconducting transition. In Fig. 1(b) the results of the single step-method (Hor's procedure) is also displayed. The effect of the two step method on the appearance of the superconductivity is clear. The resistivity measurement shows a zero resistance below about 2.7 K.  The magnetization measurement shows that these samples to be a bulk superconductor, with the shielding fraction of about 15 $\%$. The carrier concentration, {\it n}, estimated from R$_H$ at 4.2 K, is almost temperature independent and takes the value of about 1.0 x 10$^{20}$ cm$^{-3}$. It is noteworthy that the value is the same as those of the electrochemically-synthesized samples by Kriener {\it et al.}, who reported that the shielding fraction reached about 50 $\%$.\cite{r14}

\begin{figure}
\includegraphics[clip,angle=-90,width=8.0cm]{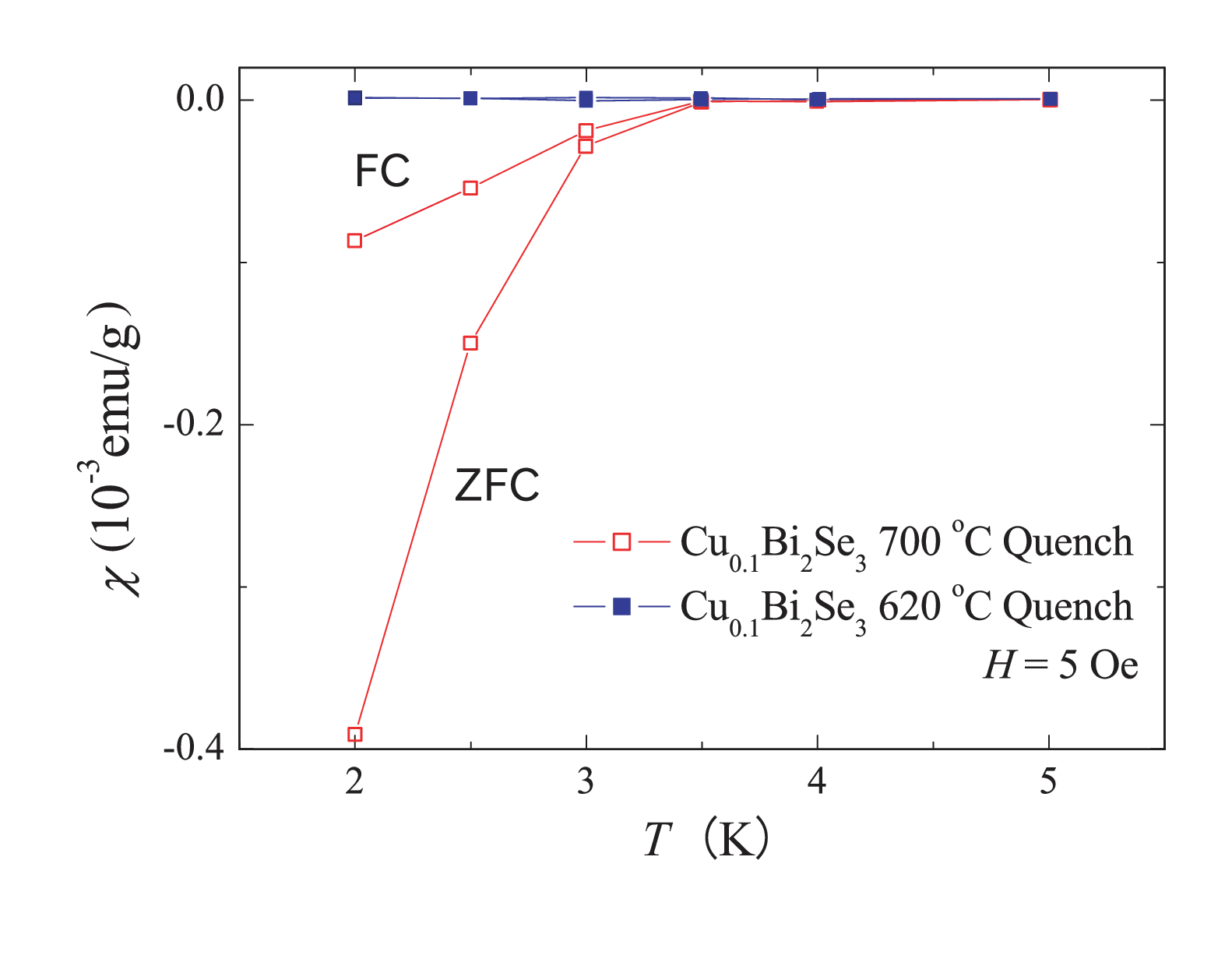}
\caption{Temperature dependence of SQUID magnetization of Cu$_{0.1}$Bi$_2$Se$_3$ samples quenched at 700 $^\circ$C (open square) and at 620 $^\circ$C (filled square). FC and ZFC represent field cooled data and zero-field cooled data, respectively.}
\label{f2}
\end{figure}

Next, we investigate the dependence of superconducting properties on the quench temperature. Thin quartz tubes (7 {\it mm$\phi$}) were used in order to measure a lump of materials synthesized in one process. Figure $\ref{f2}$ shows the temperature dependence of magnetic susceptibility for samples with x=0.1 quenched at 700 $^\circ$C and 620 $^\circ$C. The Cu composition, x=0.1, is lower than that reported by Hor {\it et al.}, where they observed superconductivity. The result clearly indicates that the appearance of the superconductivity depends on the quench temperature, and gives clear clues to establish novel reproducible synthetic method for superconducting Cu$_x$Bi$_2$Se$_3$. 

\begin{figure}
\includegraphics[clip,width=8.0cm]{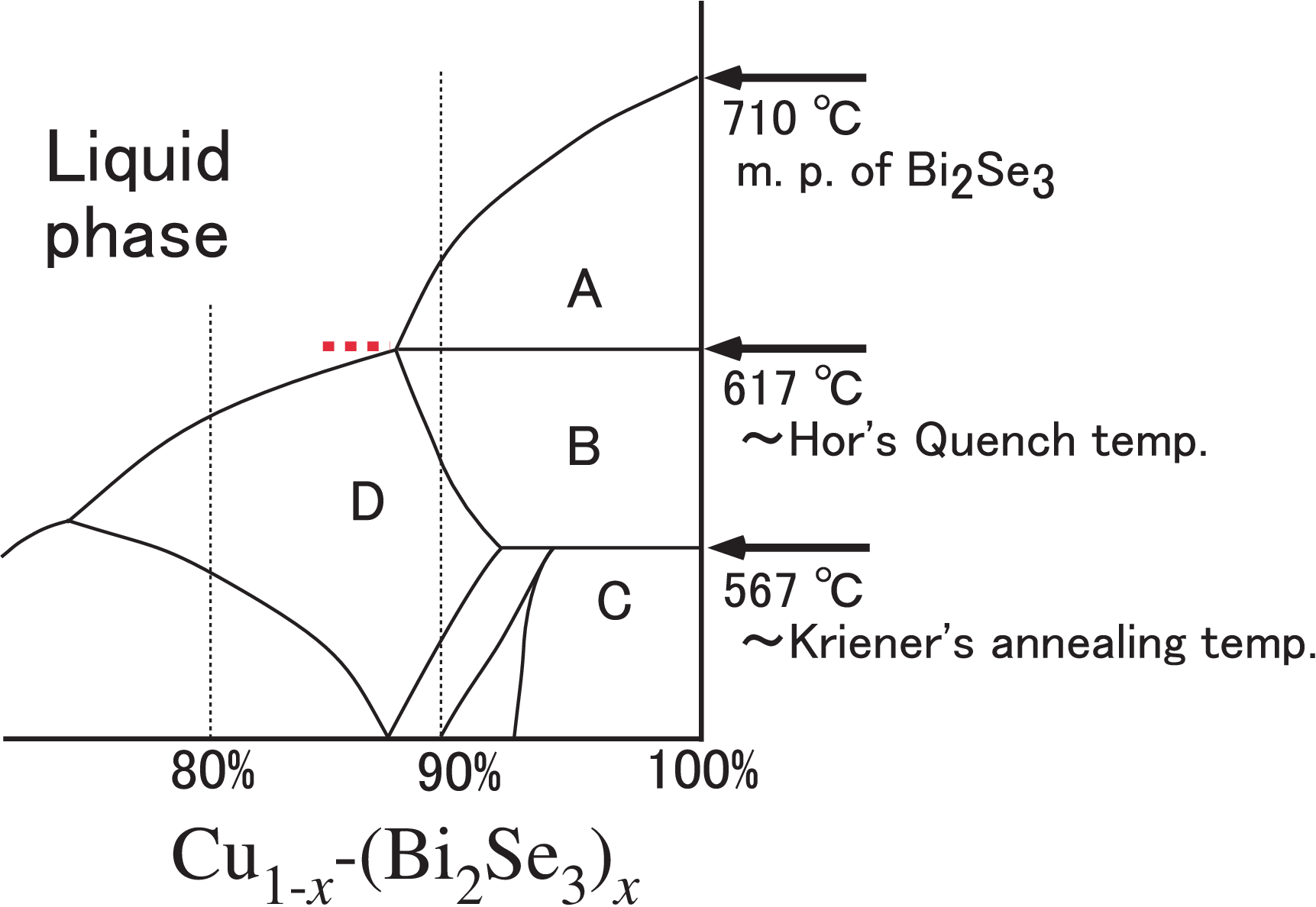}
\caption{Schematic view of Cu$_x$-Bi$_2$Se$_3$ phase diagram in the high {\it x} region (A) Liquid+Bi$_2$Se$_3$, (B) Liquid+CuBi$_3$Se$_5$+Bi$_2$Se$_3$, (C) Bi$_2$Se$_3$+CuBi$_3$Se$_5$+BiSe-based material, (D) Liquid+CuBi$_3$Se$_5$.  The optimal composition where Hor {\it et al.} found the superconductivity is indicated by the horizontal dashed line in the figure. (Cu$_{1-x}$-(Bi$_2$Se$_3$)$_x$ with 0.87$\leq${\it x}$\leq$0.89.)}
\label{f3}
\end{figure}

Recently, Babanlya {\it et al.} reported poly-thermal phase diagram of Cu-Bi$_2$Se$_3$.\cite{r20} We try to understand how superconducting samples are obtained by our two step method, based on this phase diagram. Figure $\ref{f3}$ shows the schematic view of Cu$_{1-x}$-(Bi$_2$Se$_3$)$_x$ phase diagram in the high {\it x} region. According to the phase diagram, the vertical line at Cu$_{0.1}$-(Bi$_2$Se$_3$)$_{0.9}$ crosses the phase boundary line, which divides the area above 620 $^\circ$C into the Liquid phase and the phase of Liquid+Bi$_2$Se$_3$. This should account for our previous result that the superconductivity appeared only in the sample quenched at 700 $^\circ$C. These results clearly indicate that the emergence of the superconductivity needs the quench at the liquid phase of Cu-Bi-Se mixtures.  In the sample quenched at 620 $^\circ$C, Bi$_2$Se$_3$ alone crystallizes without the intercalation of Cu atoms.  Thus, single crystals of Bi$_2$Se$_3$ naturally do not show the superconductivity. Namely, a recrystallization process to purify chemicals occurs. It should be noted that for the {\it x} range from 0.1 to 0.3 in Cu$_x$Bi$_2$Se$_3$ where Hor {\it et al.} observed superconductivity is also located in the liquid phase at 617 $^\circ$C.@

Our findings suggest the possibility that superconducting Cu-intercalated Bi$_2$Se$_3$ is produced in the wide {\it x} range, since the liquid phase occupies widely in the Cu-Bi$_2$Se$_3$ phase diagram.  In fact, by quenching the mixtures at 750 $^\circ$C, which is higher than the melting point of Bi$_2$Se$_3$, we observed the superconductivity for the {\it x} range from 0.03 to 0.5, which is rather wider range than reported by Hor {\it et al.}  However, the transition temperature and the volume fraction seemed to have no correlation with the stoichiometric Cu concentration. So, we investigate the sample produced in the same quartz tube in detail. 
 
\begin{figure}
\includegraphics[clip,angle=-90,width=8.0cm]{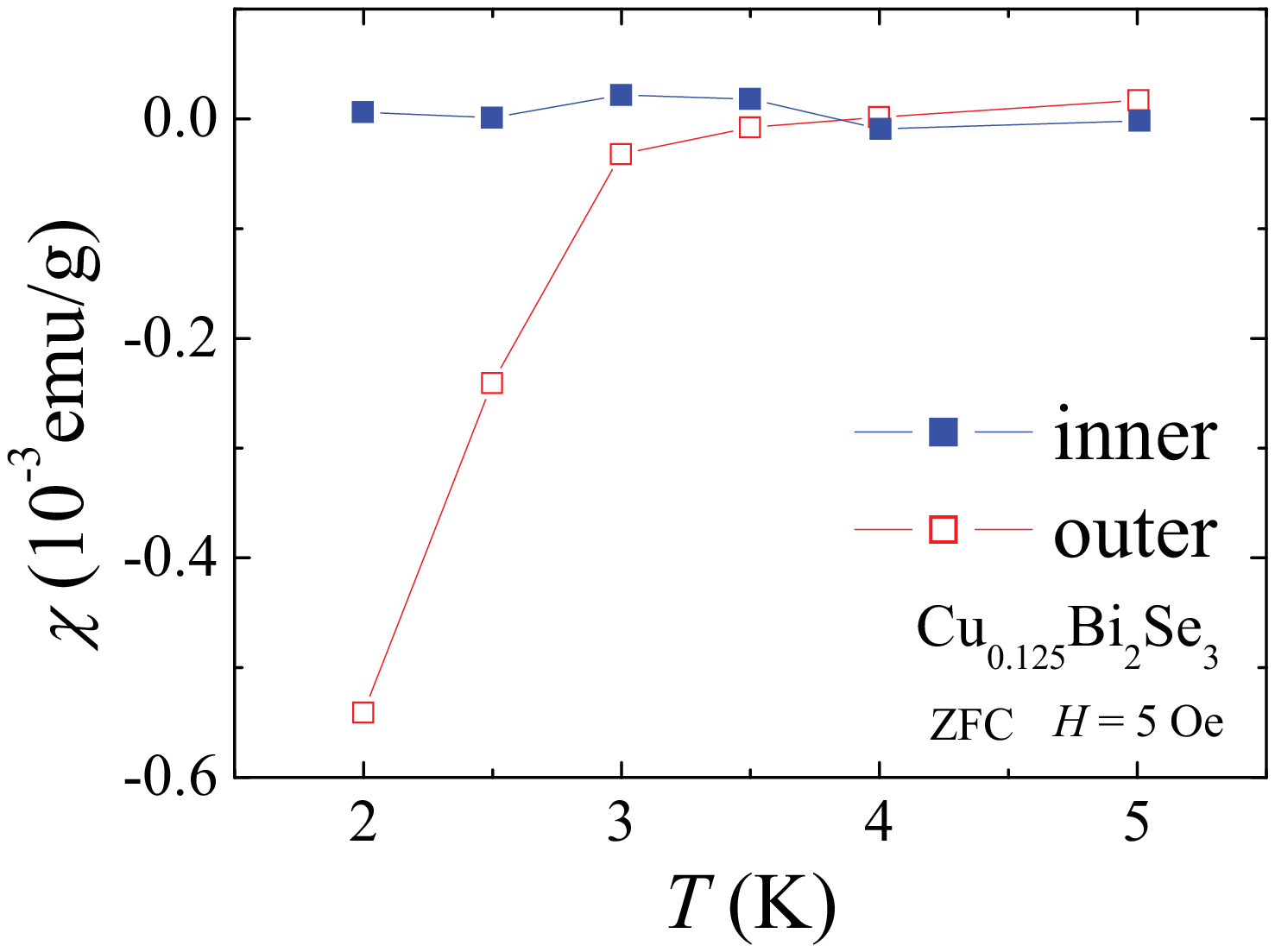}
\caption{Temperature dependence of the magnetization of Cu$_{0.125}$Bi$_2$Se$_3$ samples located at outer part (open square) and central one (filled square) in the same quartz tube.}
\label{f4}
\end{figure}

We examine the location dependence in a quartz tube for the existence of superconductivity. Figure $\ref{f4}$ shows that the temperature dependence of the magnetization of Cu$_{0.125}$Bi$_2$Se$_3$ samples, quenched at 750 $^\circ$C, located at outer part and central part in the same quartz tube. The samples were collected as powders from the outer part and the central part within 1 mm of the same cylindrically-shaped lump, the diameter of which is 8 mm. This result indicates that the outer part, rapidly-solidified by the quenching process, shows the superconductivity while the slowly-solidified central part does not show superconductivity. This suggests the possibility that the concentration of the intercalated Cu depends on the location in a cylindrically-shaped lump.

\begin{figure}
\includegraphics[clip,angle=-90,width=8.0cm]{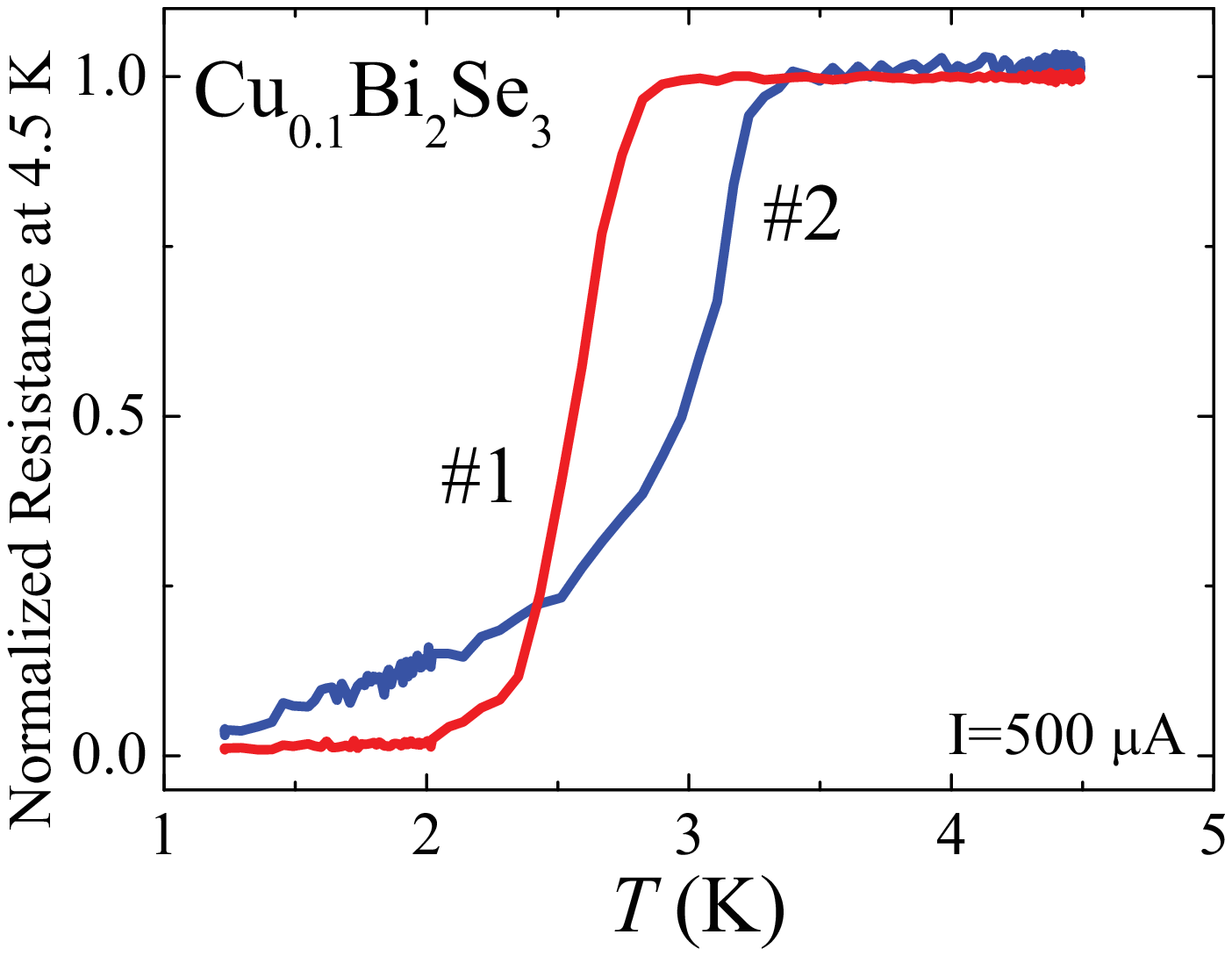}
\caption{Temperature dependence of normalized resistance at 4.5 K of samples synthesized in the same quartz tube. The difference in T$_c$ should be due to that in the Cu concentration.}
\label{f5}
\end{figure}

Figure $\ref{f5}$ shows the temperature dependence of the resistivity of two samples cut from the same ingot at around the superconducting transition. As predicted above, the samples even in the same quartz tube show different transition temperatures. This difference in the transition temperature should be due to the difference in the Cu concentration in the sample, since Kriener {\it et al.} reported the transition temperature depends on the intercalated Cu concentration.\cite{r14}These results indicate that it is difficult to synthesize homogeneously Cu-intercalated Bi$_2$Se$_3$ even by our two step method.  

It should be noted that the production of Cu intercalated Bi$_2$Se$_3$ by the melt growth method must be due to the strong stability of a Bi$_2$Se$_3$ layer unit, which forms a Bi$_2$Se$_3$ single crystal, being connected by the Van der Waals force. In fact, when a Bi$_2$-Se$_3$ mixture is quenched even at 900 $^\circ$C, much higher than the melting point of Bi$_2$Se$_3$, single crystals are formed, even though amorphous solids are mostly formed in the case of ordinary materials for such conditions. The quenching process helps Cu atoms to be left behind in between Bi$_2$Se$_3$ layer units in the single crystal, and they consequently act as electron dopants, leading to the emergence of the superconductivity.

In the phase diagram of Fig. 3, there is another phase-line at 567 $^\circ$C in the higher {\it x} region of Cu$_{1-x}$-(Bi$_2$Se$_3$)$_x$. This temperature probably coincides with the temperature where Kriener {\it et al.} annealed their samples after the electrochemically-intercalation process to emerge superconductivity.  To clarify the effect of the annealing on samples may give clues to obtaining samples with better quality.\\

In summary, we reported the reason why the conventional melt growth method for synthesizing Cu-intercalated Bi$_2$Se$_3$ could not achieve good results, and develop the improved method. Avoiding the production of Cu$_2$Se and quenching Cu-Bi-Se mixtures at the liquid phase were keys to obtaining samples with good superconducting characteristics in a reproducible manner. Investigation of the physical properties of Cu$_x$Bi$_2$Se$_3$ samples prepared by this method will contribute to the physical understanding of topological superconductivity, which is in progress.\\

This work was partially supported by Grants-in-Aid for Young Scientist (B) (No. 23740255) from the Ministry of Education, Culture, Sports, Science and Technology (MEXT), Japan.


%
%

%


\bibliography{your-bib-file}

\begin{thebibliography}{9}
\bibitem{r0} C. L. Kane and E. J. Mele, Phys. Rev. Lett. {\bf 95}, 146802 (2005).
\bibitem{r1} M. Z. Hasan and C. L. Kane, Rev. Mod. Phys. {\bf 82}, 3045 (2010).
\bibitem{r2} X.-L. Qi and S.-C. Zhang, Rev. Mod. Phys. {\bf 83}, 1057(2011).
\bibitem{r3} Y. Xia, D. Qian, D. Hsieh, L. Wray, A. Pal, A. Bansil, D. Grauer, Y. S. Hor, R. J. Cava, and M. Z. Hasan, Nature Phys. {\bf 5}, 398 (2009).
\bibitem{r4}  Y. L. Chen, J. G. Analytis, J.-H. Chu, Z. K. Liu, S.-K. Mo, X. L. Qi, H. J. Zhang, D. H. Lu, X. Dai, Z. Fang, S. C. Zhang, I. R. Fisher, Z. Hussain, and Z.-X. Shen, Science {\bf 325}, 178 (2009).
\bibitem{r4a} D. Hsieh, et al., Science {\bf 323} 919 (2009).
\bibitem{r5} Peng Cheng, {\it et al.}, Phys. Rev. Lett. {\bf 105}, 076801 (2010).
\bibitem{r6} T. Hanaguri, K. Igarashi, M. Kawamura, H. Takagi, and T. Sasagawa, Phys. Rev. B {\bf 82}, 081305(R) (2010).
\bibitem{r6k} L. Fu and C. L. Kane, Phys. Rev. Lett. {\bf 100}, 096407 (2008).
\bibitem{r7} A. P. Schnyder, S. Ryu, A. Furusaki, and A. W. W. Ludwig, AIP Conf. Proc. {\bf 1134}, 10 (2009).
\bibitem{r8} A. Kitaev, AIP Conf. Proc. {\bf 1134}, 22 (2009).
\bibitem{r12} M. Sato, Phys. Rev. B 79, 214526 (2009). 
\bibitem{r13} M. Sato, Phys. Rev. B {\bf 81}, 220504 (2010).
\bibitem{r9} F. Wilczek, Nature Phys. {\bf 5}, 614 (2009).
\bibitem{r10} Y. S. Hor, A. J. Williams, J. G. Checkelsky, P. Roushan, J. Seo, Q. Xu, H. W. Zandbergen, A. Yazdani, N. P. Ong, and R. J. Cava, Phys. Rev. Lett. {\bf 104}, 057001 (2010).
\bibitem{r11} L. A. Wray, S. Xu, Y. Xia, Y. S. Hor, D. Qian, A. V. Fedorov, H. Lin, A. Bansil, R. J. Cava, and M. Z. Hasan, Nature Phys. {\bf 6}, 855 (2010).
\bibitem{r14} M. Kriener, K. Segawa, Z. Ren, S. Sasaki, S. Wada, S. Kuwabata, and Y. Ando, Phys. Rev. B {\bf 84}, 054513 (2011). \label{r14L}
\bibitem{r15} T. Kirzhner, E. Lahoud, K. B. Chaska, Z. Salman, and A. Kanigel, Phys. Rev B {\bf 86}, 064517 (2012). 
\bibitem{r16} S. Sasaki, M. Kriener, K. Segawa, K. Yada, Y. Tanaka, M. Sato, and Y. Ando, Phys. Rev. Lett. {\bf 107}, 217001 (2011).
\bibitem{r17} N. Levy, T. Zhang, J. Ha, F. Sharifi, A. Alec Talin, Y. Kuk, and J. A. Stroscio, arXiv:1211.0267.
\bibitem{r18} H. Peng, D. De, B. Lv, F. Wei, and C.-Wu Chu, arXiv:1301.1030.
\bibitem{r19} G. P. Bernardini, A. Catani, and Mineral. Deposita., {\bf 3}, 375-380 (1968).
\bibitem{r20} N. Babanly, Yu. Yusibov, Z. Aliev, M. Babanly, Russ. J. Inorg. Chem., {\bf 55}, 1471 (2010).
\end{thebibliography}

\end{document}